\documentclass[10pt, conference, compsocconf]{IEEEtran}

\usepackage[pdftex]{graphicx}
\DeclareGraphicsExtensions{.pdf,.jpeg,.png}
\usepackage{caption}
\usepackage{subcaption}

\usepackage[cmex10]{amsmath}

\usepackage[]{algorithm2e}
\usepackage{listings}
\usepackage{tabularx}
\usepackage{tabulary}

\usepackage{multirow}

\usepackage[hyphens]{url}

\begin{document}
%
\title{A Novel Control-flow based Intrusion Detection Technique for Big Data Systems}


\author{\IEEEauthorblockN{Santosh Aditham}
\IEEEauthorblockA{Dept of Computer Science and Engineering\\
University of South Florida\\
Tampa, USA.}
\and
\IEEEauthorblockN{Nagarajan Ranganathan}
\IEEEauthorblockA{Dept of Computer Science and Engineering\\
University of South Florida\\
Tampa, USA.}
}


\maketitle

\begin{abstract}
\textit{
Security and distributed infrastructure are two of the most common requirements for big data software. But the security features of the big data platforms are still premature. It is critical to identify, modify, test and execute some of the existing security mechanisms before using them in the big data world. In this paper, we propose a novel intrusion detection technique that understands and works according to the needs of big data systems. Our proposed technique identifies program level anomalies using two methods - a \textbf{profiling method} that models application behavior by creating process signatures from control-flow graphs; and a \textbf{matching method} that checks for coherence among the replica nodes of a big data system by matching the process signatures. The profiling method creates a process signature by reducing the control-flow graph of a process to a set of minimum spanning trees and then creates a hash of that set. The matching method first checks for similarity in process behavior by matching the received process signature with the local signature and then shares the result with all replica datanodes for consensus. Experimental results show only 0.8\% overhead due to the proposed technique when tested on the hadoop map-reduce examples in real-time. 
}
\end{abstract}


\section{Introduction}

The architectures for big data systems rely on parallel execution techniques like mapreduce \cite{:mapred} for fast processing. With the growing popularity of real-time data processing in big data environments, there is a pressing need to re-imagine the traditional computing techniques. For example, data locality in popular big data system distributions like hadoop \cite{:hadoop} and spark \cite{:spark} is redefined as bringing compute to data instead of the traditional approach of the moving the data that needs to get processed. This trend of re-inventing the traditional methods do not necessarily transform to the security needs of big data. The security features implemented in big data systems are still based on traditional methods for systems based on general purpose machines. User authentication, multi-level data access control and logging are typically used for security in big data \cite{:kerberos}. Data encryption is slowly being adopted in the big data field, but it is limited by big data properties like volume and velocity. As we covered in our previous work \cite{:santosh}, big data security is premature and there is a lot of scope for improvement in this area. For instance, the current security standards for big data systems assume system-level consistency which is not necessarily true always. We demonstrated in our previous work \cite{:santosh} that big data platforms can be affected by insider attacks. In this work, we concentrate on detecting process-level intrusions within big data systems.

\begin{figure}
\centering
\includegraphics[width=0.45\textwidth]{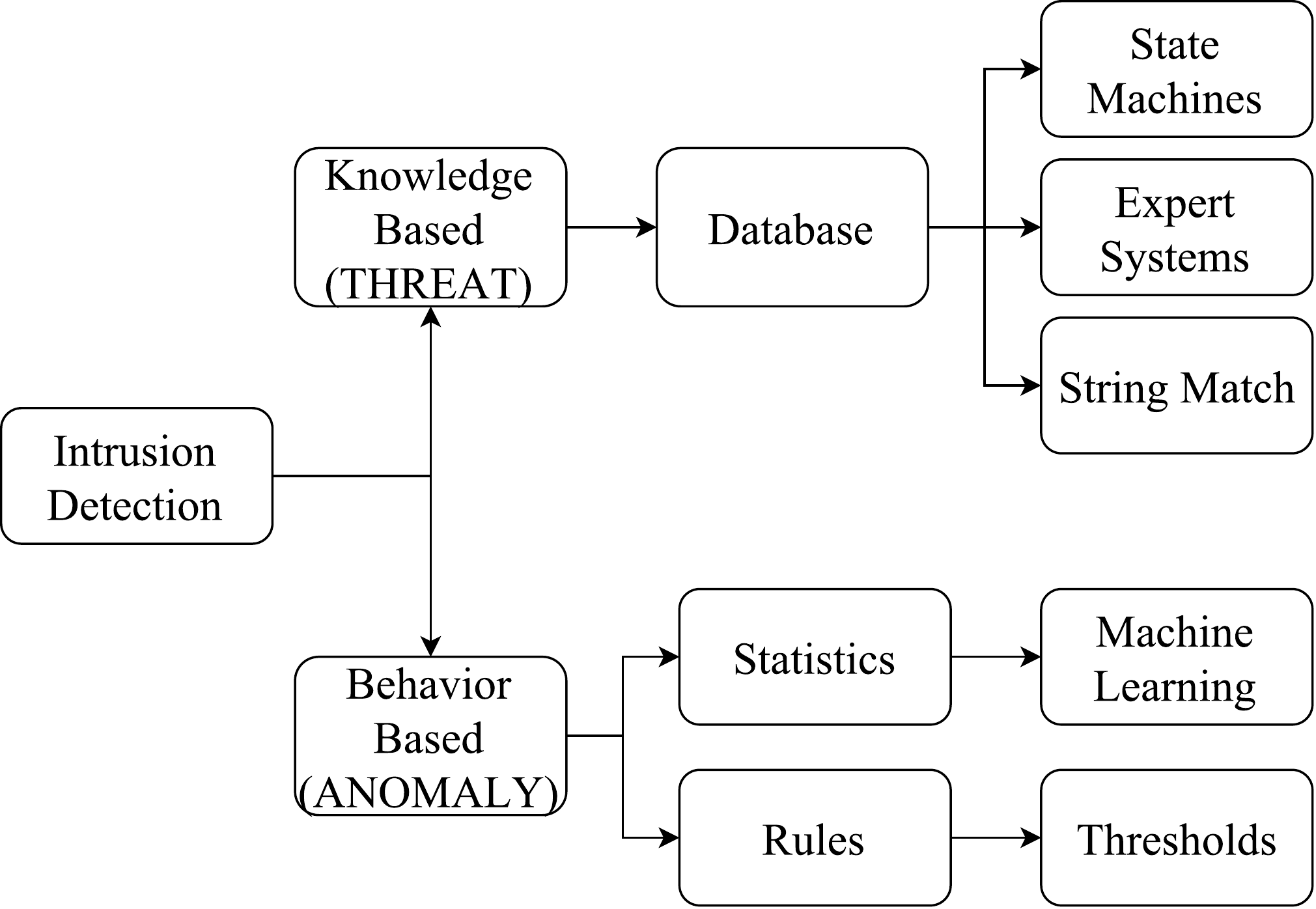}
\caption{A taxonomy of Intrusion Detection Techniques}
\label{fig_tax}
\end{figure}

Intrusion detection systems (IDS) can identify malicious use based on their knowledge of possible threats or by learning from the behavior of programs. Knowledge-based IDS usually search a program for known threat signatures that are stored in a database. With new and zero-day attacks emerging regularly, it is impractical to have a pre-populated database of all possible threats. Even if it is assumed to have such a database, maintaining it would require a lot of resources and running search queries against it would be expensive. Behavior based IDS tries to model, analyze and compare application behavior to identify anomalies. This technique needs more resources and is more complex than signature-based IDS but it is more effective in a dynamically changing threat environment. Behavior based IDS generally use statistics and rules to detect anomalies. Figure \ref{fig_tax} gives a taxonomy of the different types of IDS.

In today's internet age, a distributed implementation of IDS is needed for which aggregation, communication and cooperation are key factors of success. Distributed IDS gives centralized control and detects behavioral patterns even in large networks but it has to be employed at multiple levels: host, network and data \cite{:cids}. Hence, using big data in general-purpose distributed IDS implementations is recommended for faster processing. In this work, we concentrate on IDS that can be used for security within big data systems. IDS within a big data system favors anamoly-based IDS when compared to knowledge-based IDS because of the naturally large and ever increasing scope of threats. 

Using control-flow graphs for logic level intrusion detection is a commonly known idea \cite{:cfg-match1, :cfg-match2, :graph-compare-exe}. For example, control-flow integrity \cite{:ligatti} is a security mechanism that can identify misuse of application logic bugs, like buffer-overflow attacks. Though CFGs are generally sparse graphs, they can grow very big in size. Hence, it is important to design IDS techniques that can work with a reduced representation of CFGs. A Minimum Spanning Tree (MST) contains all vertices and only some paths of its source graph and the number of MSTs for sparse graphs is generally less. Hence, a set of MSTs extracted from a CFG can be used for IDS that detects program level anomalies.  

In this paper, we propose a control-flow based intrusion detection technique for big data systems. The proposed technique checks for program level anomalies in big data applications by analyzing and comparing the control-flow behavior of all processes running inside a big data system. The proposed intrusion detection technique is divided into two parts. First, the control-flow of each process running on a data node in the big data cluster is locally analyzed. This is done by extracting a set of MSTs from the instruction level CFG of a compiled program. The extracted set of MSTs are hashed and stored in an array called the \textit{program signature}. Then, the stored program signature is encrypted and shared with other replica nodes that run the same program. In the second step, the received encrypted program signature is decrypted and matched with the local version to check for coherence. Matching two program signatures involves finding a perfect match for every MST in a signature within the set of MSTs of the other. The result of the matching step is then shared with replica nodes for consensus. Our technique is designed to be simple, scalable and efficient in identifying both control-flow and brute-force attacks.

The rest of this paper is organized as follows. Section II gives some background about big data systems, control-flow graphs and IDS. The various related works are also discussed here. Section III explains the proposed intrusion detection technique in detail. Experimental setup and results are thoroughly discussed in Section IV. Finally, Section V gives the conclusion and future work.

\section{Background and Related Work}
In this section, background about the three topics - big data systems, control-flow graphs and intrusion detection is provided. The related works are briefly outlined here.

\subsection{Big Data Systems}
Big data systems are data driven and their work can be classified into 2 major tasks - writing user data to the disk for storage and; reading stored data when user requests for it. Typically, this data is quantified in units called \textit{blocks}. For fast and fault-tolerant service, big data systems rely on replication of data blocks which in turn demands data consistency. Big data systems cannot afford to have read or write service-level inconsistency. The motivation for this work comes from a weak assumption in the big data community that the services used by a big data system to maintain data consistency are never attacked. It is our knowledge that this problem has not been widely addressed before.

To propose an IDS for big data services, it is important to understand how the services work. For this, we picked 2 popular big data services - reads and writes. When a client (or user) wants to write a block, the namenode picks \textit{n} data nodes from the big data cluster to complete this task where \textit{n} is the replication factor of the cluster. First the namenode checks if the datanodes are ready. It sends a ready request to datanode1 which when ready, forwards that request to datanode2  and so on. When the namenode knows that all \textit{n} datanodes are ready, it asks the client to start writing. The client only writes to datanode1 which is subsequently written on to datanode2, datanode3 and so on. In case of any failure, namenode orders a new datanode to maintain block replicas. When the client wants to read a block, namenode gives the client a list of all datanodes that have the block and the client picks first datanode. If there is a problem reading from datanode1, the client request gets forwarded to the next datanode that has a copy of the same block. 

\subsection{Control-flow Graphs}


\begin{figure}[t!]
    \centering
    \begin{subfigure}[b]{0.45\textwidth}
        \centering
        \includegraphics[height=2in]{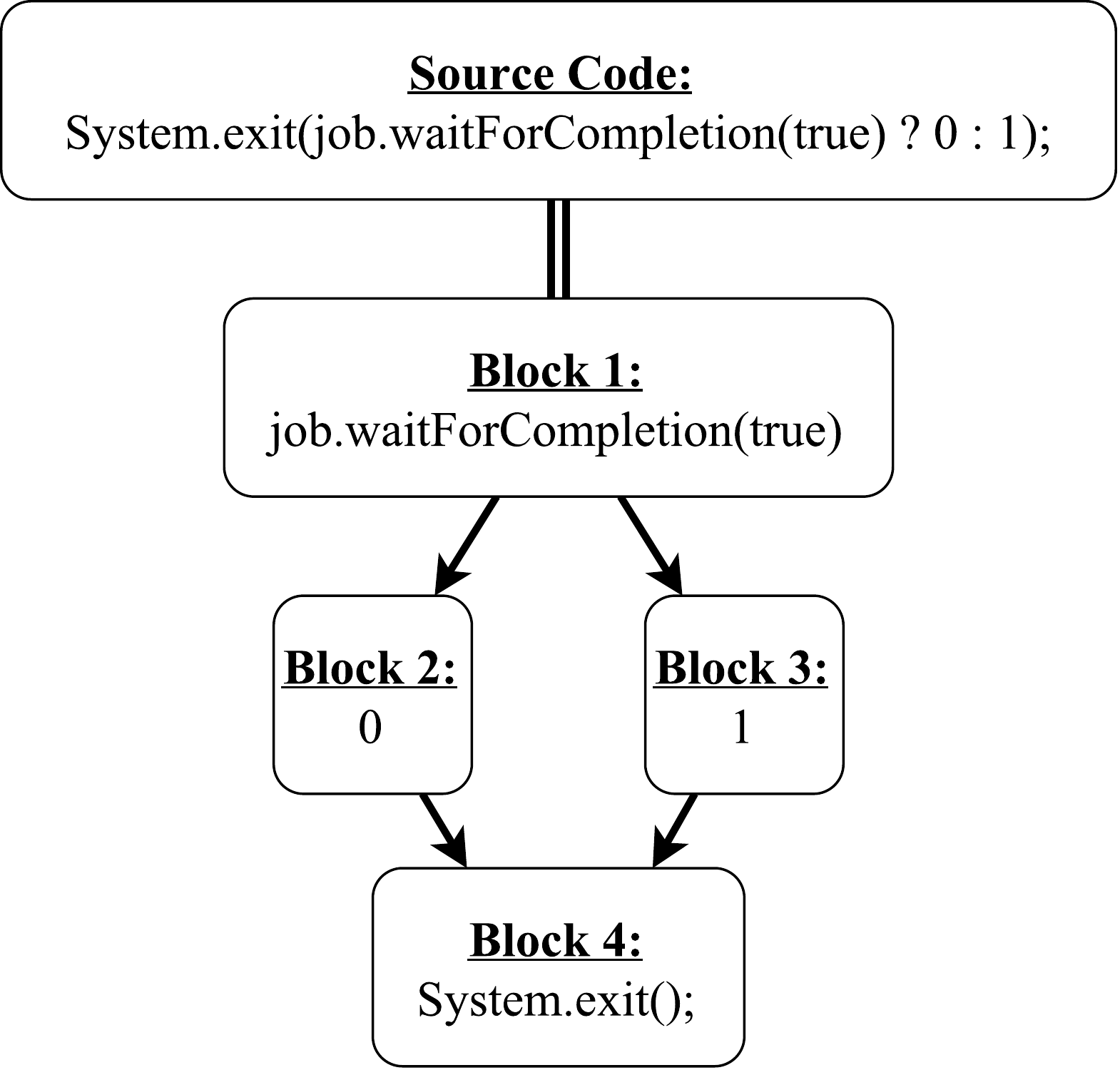} 
        \caption{Source code to basic blocks}
    \end{subfigure}
    ~ 
    \begin{subfigure}[b]{0.45\textwidth}
        \centering
        \includegraphics[height=1in]{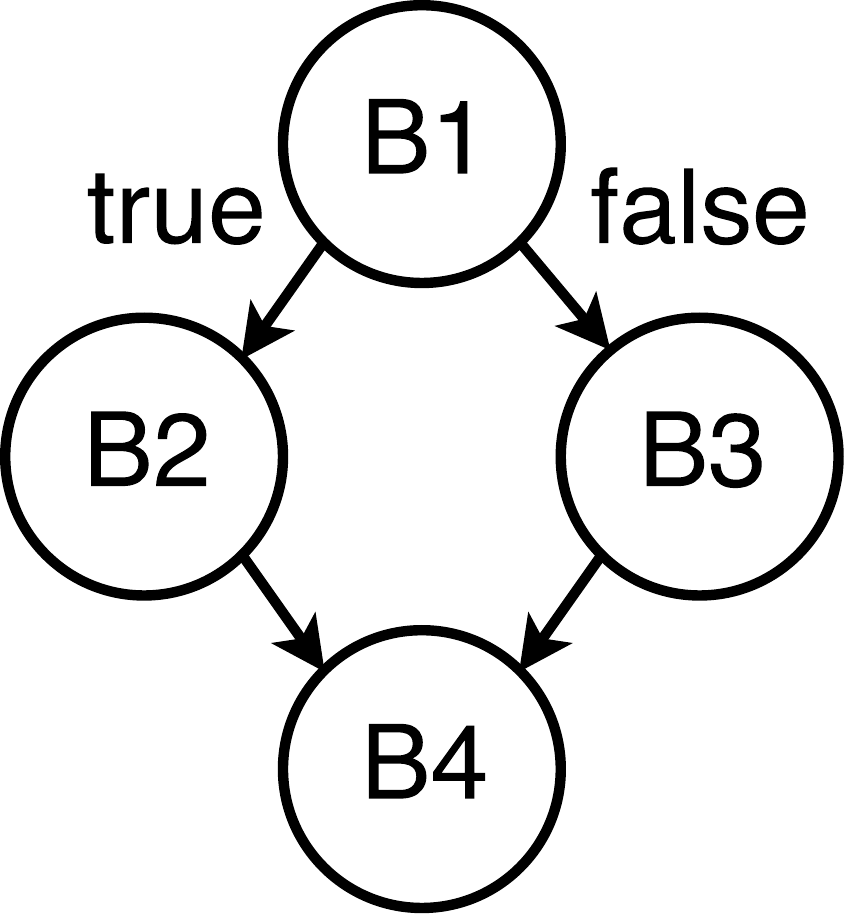} 
        \caption{Basic-block CFG}
    \end{subfigure}
    ~ 
    \begin{subfigure}[b]{0.2\textwidth}
        \centering
        \includegraphics[height=1in]{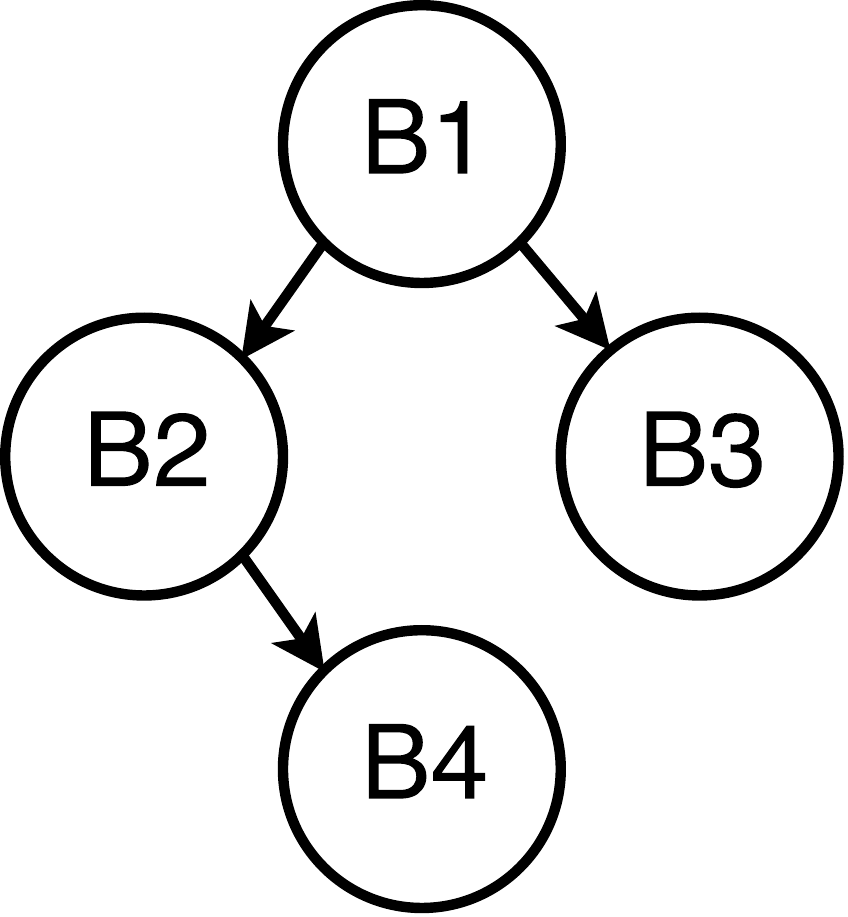}
        \caption{MSA of CFG}
    \end{subfigure} 
    \begin{subfigure}[b]{0.2\textwidth}
        \centering
        \includegraphics[height=1in]{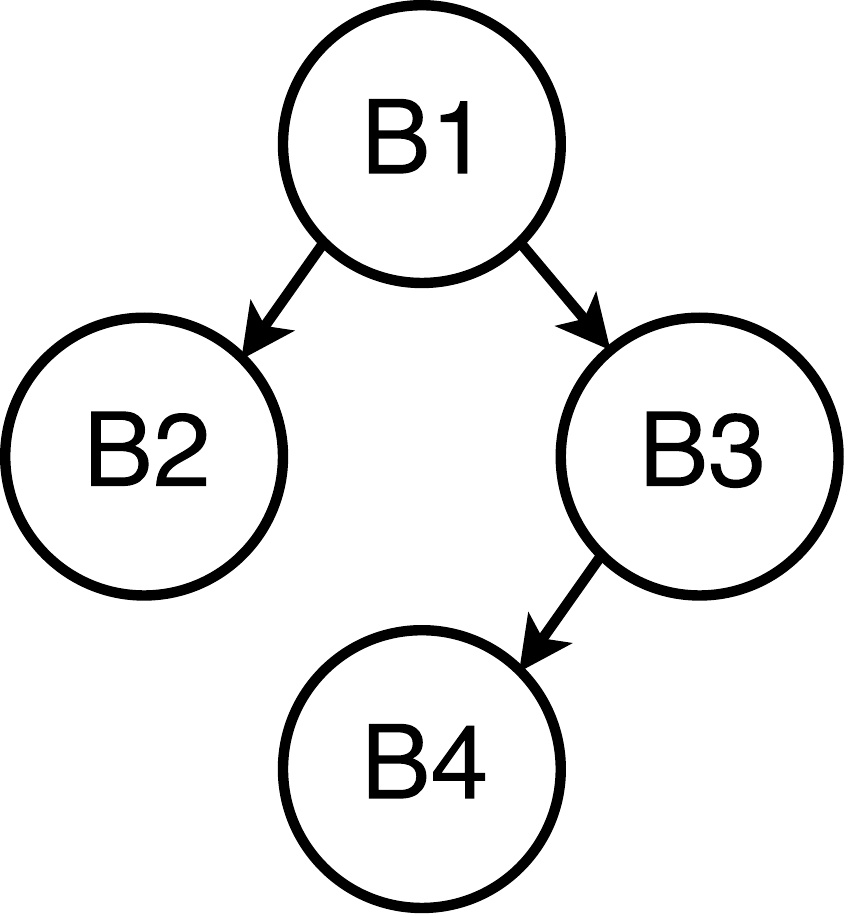}
        \caption{Another MSA of CFG}
    \end{subfigure}    
    \caption{Multiple MSAs of same CFG}
\label{fig_ex1}
\end{figure}

A control-flow graph (CFG) is a directed graph representation of a program and usually a sparse graph. CFGs include all possible control paths in a program. This makes CFG a great tool to obtain control-flow behavior of its process. Vertices in a CFG give the level of detail, such as instruction-level or basic block level, that cannot be further divided. Edges in CFG represent control jumps and are classified into two types - forward and backward. Branch instructions, function calls, conditional and unconditional jumps account for forward edges. Virtual calls and indirect function calls are also considered as forward edges but their destinations are difficult to determine. Loops and returns generally account for backward edges. The integrity among duplicate processes that run on replica nodes of a big data system can be verified with the information available in a CFG \cite{:cfg-correct}. Similarity check between program logic of two programs can be performed by comparing their CFGs for isomorphism. There are many ways to check for such graph isomorphism \cite{:algos-graphsim, :isomorphism} but analyzing the similarity of two processes by conducting CFG level graph isomorphism is hard and time consuming. Graph isomorphism is a complex problem, sometimes known to be NP-complete as well \cite{:cfg-match2}. To reduce the complexity of graph algorithms, CFGs can be reduced to trees or subgraphs before performing any coherence or integrity checks \cite{:cfg-reduce}. A CFG can be converted to a tree using methods such as Depth-first traversal. Several tree structures like Dominator Tree, Minimumm Spanning Tree (MST), Minimumm Spanning Arborescence (MSA) can be extracted form CFGs \cite{:edmonds1, :edmonds2, :edmonds3}. For this work, MST and MSA can be used interchangeably. CFGs can be broken into subgraphs using methods like k sub-graph matching and graph coloring.  Some popular methods for graph reduction and graph comparison that can be found in the literature are given below (assume graphs to have n vertices and m edges):

\begin{itemize}
\item \textit{Based on Edit Distance}: Using Smith-Waterman algorithm with Levenshtein distance to identify similarity between two graphs represented as strings \cite{:edit-distance}. The time complexity is O(nm).
\item \textit{Based on Traversal}: (a) A preorder traversal of a graph G where each node is processed before its descendants. (b) A reverse postorder in a DAG  gives a topological order of the nodes \cite{:traversal}.
\item \textit{Based on Dominator trees}: A data structure built using Depth First Search or using the method proposed by Tarjan in \cite{:tarjan1}. Tarjan's method has a time complexity of O((n+m)log(n+m)).
\item \textit{Based on Reachability}: Transitive reduction of a sparse graph to another graph with fewer edges but same transitive closure \cite{:tarjan2}. The time complexity is O(nm).
\end{itemize}  

In this work, we chose to reduce a CFG to a set of MSTs because CFGs are generally sparse graphs and hence the size of the set of MSTs will be finite and small. Edmond's algorithm can be used to extract MSTs from a digraph \cite{:edmonds1, :edmonds2, :edmonds3}. Since an MST contains all vertices of its graph, there will be no loss in the program instruction data. Depending on the connectedness of the graph, the edge count will defer between the CFG and MST representation of a program. Figure \ref{fig_ex1} shows transformation of a line of java code to basic blocks of bytecode to CFG to set of MSAs. Vertices B1, B2, B3, B4 are the basic blocks formed from java bytecode. There exists an O(m + n log n) time algorithm to compute a min-cost arborescence \cite{:edmonds1}. Alternately, another approach for converting a CFG to MST using union find is used by popular compilers like llvm and gcc for security purposes \cite{•}. One known disadvantage of using CFGs and MSTs for security is that dynamic link library calls cannot be verified.  
          
\subsection{Intrusion Detection Systems}
Traditionally, IDS checks for known malware in programs by performing signature matching on a threat database \cite{:ids-book}. Signature match using exact string matching is limited in its scope. This is because variants of same attack will have different signatures. Recently, methods to detect new malwares using statistical machine learning have been proposed. Static analysis using CFG is another efficient way to detect intrusions but it is very complex \cite{:static-ids}. Converting a CFG to a string and implementing string matching is another way to deal with this problem but the solution will not be polynomial. Also, CFG at basic block level can have basic block variants that look different but perform the same function. To deal with these shortcomings, many approximate matching techniques have been proposed. Tracing applications to get their CFG is another approach that is used in applications like xtrace, pivottrace etc \cite{:hdfs-tracing, :pivot-tracing}. In case of big data systems, data nodes usually have the same processor architecture. Hence it can be assumed that there will be no variants when the CFG is constructed at byte-level. It is then sufficient to verify similarity among the CFGs of two processes to confirm coherence in the nodes of a big data system.

\begin{figure*}[t!]
    \centering
    \begin{subfigure}[b]{1\textwidth}
        \centering
        \includegraphics[height=0.5in, width=0.8\textwidth]{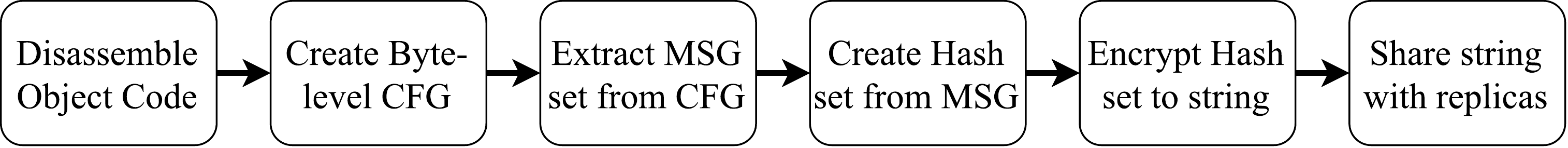}
		\caption{Steps in Generating Process Signatures}
    \end{subfigure}
    ~ 
    \begin{subfigure}[b]{1\textwidth}
        \centering
        \includegraphics[height=0.5in, width=0.8\textwidth]{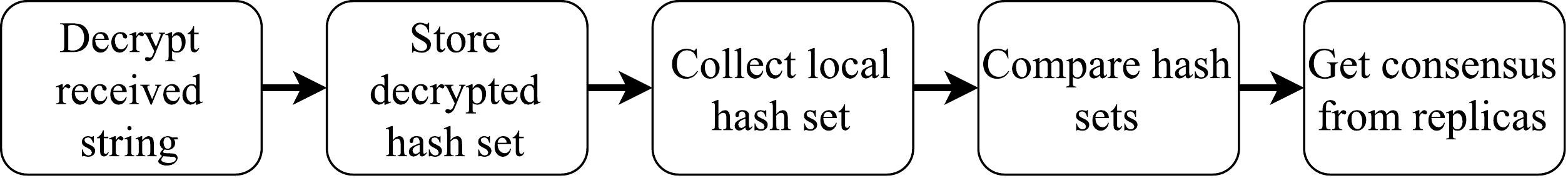}
		\caption{Steps in Matching Process Signatures}
    \end{subfigure}
\caption{Proposed Algorithm for Intrusion Detection}
\label{fig_algo}
\end{figure*}

\section{Proposed Technique}
In this section, we describe our proposed two-step intrusion detection technique for big data systems. The first step involves capturing the control-flow of a process running on a datanode of the big data system. The second step involves process-level similarity check followed by consensus among replica datanodes.

\subsection{Generating Process Signatures} 

In this work, we emphasize on process level intrusion detection by observing coherence in the behavior of duplicate processes running on replica datanodes of a distributed big data system. To capture the program behavior, the first step is to identify a representation of the program that has the information we need and filters out all other data. We call this representation as the program signature. Since our goal is to identify intrusions from control-flow mismatch, our program signatures should contain all possible control flow information of a program.

Compiled source code of a program is generally used to generate static CFG. Since most big data frameworks use a virtual machine (like JVM), an instruction level CFG in this context is generated from java byte code. In this work, disassembled object code (DOC) from java byte code is used as input to generate the CFG at instruction level. It is important for the program signature to contain only the information that is necessary. Hence, every CFG is converted into a set of MSTs that are later used to generate the program signature. In this work, we propose the idea of representing a program by a set of MSTs/MSAs that can be extracted from a byte-level CFG using Edmonds algorithm. This set of MSTs that are extracted from a CFG are further filtered to only the set of edge-disjoint MSTs. There are many versions proposed for Edmonds algorithm \cite{:edmonds1, :edmonds2, :edmonds3} and for this work we used a version from NetworkX graph library \cite{:nx} that generates edge disjoint spanning trees from the root vertex of a given digraph. Once a minimal representation of the logic in a program is obtained in the form of an MSA, it is converted into a string by listing the node list first followed by edge list, which is in accordance to the DOT format representation. 

The length of a MST string in DOT format is dependent on program size. To make the comparison step faster, we convert the variable length MST strings of a program to fixed length strings using hashing. The extracted set of edge-disjoint MSTs are hashed using popular hashing algorithms like SHA or MD5 to generate a set of fixed-length hash strings. Since a sparse graph like CFG can have multiple MSAs, the program signature can be a single hash string or a set of hash strings. Having all possible MSAs in the program signature makes the graph similarity check more reliable. In the end, a \textit{program signature} is a set of fixed-length strings.

Program signatures are encrypted before being shared with replica datanodes for tighter security. The private key for encryption is generated from a harcoded master key if we use secure hardware like the one proposed in our previous work \cite{:santosh}. Every datanode in a big data system runs the proposed \textit{profiling method} for every running process and it includes all the steps involved in converting the compiled binary of a program to its program signature. A pictorial representation of the steps in profiling method is given in Figure \ref{fig_algo}.

\subsection{Matching Process Signatures}

Replication property of big data systems opens scope for new methods of implementing application logic level IDS techniques. Process similarity check among duplicate nodes of the cluster helps in checking for coherence among the replica datanodes while performing a write or read operation. When a process is scheduled to run on a datanode that hosts the primary copy of a data, a signature for that process is created by the \textit{profiling method} (Step 1) of our proposed IDS technique and that signature string is shared with all replica datanodes. In the \textit{matching method} (Step 2), these signatures received from other datanodes are decrypted and matched with the local versions of the same process. The results are shared with all other replica datanodes for consensus. For secure communication among datanodes, we intend to use the same secure communication protocol that was proposed in our previous work \cite{:santosh}. 

The most important part of the matching method is to check for similarity (or dissimilarity) between two program signatures. Generally, graph similarity check can be performed by checking node similarity and edge similarity. The following points are considered while comparing MSTs to check for similarity among programs: 
\begin{itemize}
\item MSTs are sparse graphs obtained from byte-level CFGs. Hence, checking for path sensitivity is not exponential.
\item All edges are assumed to have the same weight of 1. 
\item The total number of MSTs for a CFG is limited (by Cayley's formula \cite{:cayley}). 
\item By Edmond’s theorem, a graph which is k-connected always has k edge-disjoint arborescences.
\item Two MSTs are a perfect match if their node sets and edge sets match exactly. 
\item If edge set of one MST is a subset of the edge set of another MST, the source graphs of these MSTs are not similar.
\item Two graphs are similar if for every MST in one graph there exists a perfect match in the set of MSTs of the other graph.
\item Hashing algorithms like SHA1 or MD5 are quick and efficient.
\end{itemize}


Based on the points listed above, the following method is developed for graph similarity check. Let us consider 2 control-flow graphs G1 and G2. Let $\textless N1, E1\textgreater$  represent G1 where N1 is the node set of the graph G1 and E1 is the edge set of the graph. Similarly, $\textless N2, E2\textgreater$ represents G2 where N2 is the node set of the graph G1 and E2 is the edge set of the graph. After employing a variation of Edmonds algorithm on these CFGs (such as finding all edge-disjoint MSTs), lets us assume that M1 $[ \textless N1, E1' \textgreater ]$  is the set of MST/MSA for G1 and M2 $[ \textless N2, E2' \textgreater ]$ is the set of MST/MSA for G2. In order to check for similarity in both graphs G1 and G2, we check if there is a perfect match in M2 for all MSTs in M1. In order to simplify the match function, we propose using a hash function on M1 and M2 that creates a unique hash for every MST. Let H1 be a set of hashes generated from M1 and H2 be the set of hashes from M2. If any hash in H1 does not exist in H2, we deduce that the graphs are not equal.

\section{Experimental Results}
In this section, the experimental setup and experiments used for testing the proposed technique are provided. The results and some analysis are also provided.

%
%

\subsection{Setup}
An Amazon EC2 \cite{:amazon} m4.xlarge instance running Ubuntu 14.04 is used to generate MSTs (and their hashes) from CFGs using SageMath. The proposed technique was implemented and tested on an Amazon EC2 big data cluster of 5 t2.micro nodes - 1 master node, 1 secondary master node and 3 datanodes with a replication factor of 3. The list of softwares used in conducting our experiments are:
\begin{itemize}
\item \textbf{SageMath} \cite{:sage} is a free open-source mathematics software system for mathematical calculations. 
\item \textbf{GraphML} \cite{:gml} is a popular graph representation format which can used to represent both CFG and MST. 
\item \textbf{Graphviz} \cite{:gv} is open source graph visualization software that takes input in DOT format and makes diagrams in useful formats. 
\item \textbf{NetworkX} \cite{:nx} is a Python language software package that provides graph algorithms like Edmonds and VF2. 
\item \textbf{Control-flow graph factory} \cite{:cfgf} is a software that generates CFGs from java bytecode (class file) and exports them to GraphML or DOT formats.
\end{itemize}

\begin{table}
  \caption{List of Hadoop Map Reduce Examples}
  \label{table_examples_hadoop}
  \centering
  \begin{tabulary}{0.48\textwidth}{| C | C | L |} \hline
    \textbf{E.No}&\textbf{Name}&\textbf{Description} \\\hline
	1&  wordmean & 				A map/reduce program that counts the average length of the words in the input files. \\\hline
	2&  pentomino & 			A map/reduce tile laying program to find solutions to pentomino problems. \\\hline
	3&  distbbp & 				A map/reduce program that uses a BBP type formula to compute the exact bits of pi. \\\hline
	4&  aggregatewordcount & 	An Aggregate based map/reduce program that counts the words in the input files. \\\hline
	5&  secondarysort & 		An example defining a secondary sort to the reduce. \\\hline
	6&  aggregatewordhist & 	An Aggregate based map/reduce program that computes the histogram of the words in the input files. \\\hline
	7&  randomwriter & 			A map/reduce program that writes 10 GB of random data per node. \\\hline
	8&  teravalidate & 			Check the results of the terasort. \\\hline
	9&  qmc &  					A map/reduce program that estimates the value of Pi using a quasi-Monte Carlo (qMC) method.\\\hline
	10& wordstandarddeviation & A map/reduce program that counts the standard deviation of the length of the words in the input files. \\\hline
	11& wordmedian & 			A map/reduce program that counts the median length of the words in the input files. \\\hline
	12& bbp & 					A map/reduce program that uses Bailey Borwein Plouffe to compute the exact digits of pi. \\\hline
	13& teragen & 				Generate data for the terasort. \\\hline
	14& sudoku & 				A Sudoku solver. \\\hline
	15& wordcount & 			A map/reduce program that counts the words in the input files. \\\hline
	16& multifilewc & 			A job that counts words from several files. \\\hline
  \end{tabulary}
\end{table}

\subsection{Experiments}
The proposed intrusion detection technique was tested using 16 hadoop map-reduce examples that can be found in all hadoop distributions. These examples cover a wide range of big data applications as listed in Table \ref{table_examples_hadoop}. The class files of these examples are readily available in the hadoop distributions. First, control-flow graph factory \cite{:cfgf} was used to generate control flow graphs from the class files. These graphs are stored in graphml format and given as input to a simple SageMath \cite{:sage} script that uses NetworkX library \cite{:nx} and computes the edge-disjoint MSAs and hashes them using MD5. A C++ application was used to implement encryption and secure communication needed for the proposed IDS technique. The implementation was based on framework from \cite{:santosh}. The hashes are fixed length strings and so we restrained to using a basic numeric key based left/right shift for encryption/decryption of messages. Since there are no benchmarks for some of these examples, we executed them with minimum input requirements.         

\begin{table*}[!t]
  \caption{Hadoop Map Reduce Examples - Program level time metrics in seconds}
  \label{table_hadoop_values}
  \centering
  \begin{tabular}{|c|c|m{1cm}|m{1cm}|m{1.1cm}|m{1.3cm}|m{1cm}|m{1.1cm}|m{1cm}|m{1cm}|m{1cm}|}
    \hline
\textbf{E.No}&	  \textbf{Example}&	     \textbf{Profiling method}&	\textbf{CFG to MSA set}&	  \textbf{Hashing}&      \textbf{Matching  method}& \textbf{Avg Hash Match}&	  \textbf{Consensus}&	 	\textbf{Proposed}&	  \textbf{Exec Time}&	  \textbf{\% Time}	  	   \\\hline

1&	   	wordmean&	    		0.0216& 		0.0216&	7.89E-05&	0.0190&	0.0002&	0.0187& 	0.0407&	6.988	&0.58\%	   \\\hline
2&	   	pentomino&	   			0.0288& 		0.0288&	8.70E-05&	0.0196&	0.0013&	0.0182& 	0.0485&	4.914	&0.99\%	 \\\hline  
3&	   	distbbp*&	 			0.0567& 		0.0567&	6.29E-05&	0.0150&	0.0019&	0.0130& 	0.0718&	28.58 	&0.25\%	\\\hline   
4&	   	aggregatewordcount&	  	0.0070& 		0.007&	5.70E-05&	0.0145&	0.0002&	0.0143& 	0.0215&	19.002 	&0.11\%	 \\\hline  
5&	   	secondarysort*&	  		0.0199& 		0.0199&	5.10E-05&	0.0072&	0.0018&	0.0054& 	0.0272&	11.657	&0.23\%	  \\\hline 
6&	   	aggregatewordhist&	   	0.0066& 		0.0066&	4.20E-05&	0.0135&	0.0012&	0.0122& 	0.0201&	18.024	&0.11\%	 \\\hline  
7&	   	randomwriter&	    	0.2561&			0.2561&	8.58E-05&	0.0217&	0.0025&	0.0191& 	0.2779&	29.111	&0.95\%	  \\\hline 
8&	   	teravalidate&	    	0.0181& 		0.0181&	5.20E-05&	0.0169&	0.0001&	0.0168& 	0.0351&	5.958	&0.59\%	 \\\hline  
9&	   	qmc*&	    			0.0238& 		0.0238&	7.39E-05&	0.0202&	0.0015&	0.0186& 	0.0440&	11.657	&0.38\%	\\\hline   
10&	  	wordstandarddeviation&	0.0193& 		0.0193&	7.89E-05&	0.0098&	0.0021&	0.0076& 	0.0292&	7.112	&0.41\%	\\\hline   
11&	  	wordmedian&	  			0.0312& 		0.0312&	6.20E-05&	0.0208&	0.0020&	0.0187& 	0.0520&	7.028 	&0.73\%	\\\hline   
12&	  	bbp&	 				0.0415& 		0.0415&	9.08E-05&	0.0118&	0.0003&	0.0115& 	0.0534&	6.865 	&0.78\%	\\\hline   
13&	  	teragen&	 			0.0169& 		0.0169&	5.51E-05&	0.0131&	0.0023&	0.0108& 	0.0301&	4.905 	&0.61\%	  \\\hline 
14&	  	sudoku*&	 			0.0177& 		0.0177&	5.60E-05&	0.0156&	0.0006&	0.0150& 	0.0334&	11.657 	&0.29\%	  \\\hline 
15&	  	wordcount&	   			0.3672& 		0.3672&	6.99E-05&	0.0221&	0.0023&	0.0197& 	0.3893&	7.034 	&5.54\%	 \\\hline  
16&	  	multifilewc&	 		0.0159& 		0.0159&	5.20E-05&	0.0118&	0.0001&	0.0116& 	0.0277&	5.963 	&0.47\%	 \\\hline   \hline
\multicolumn{2}{|c|}{\textbf{Average Values}}&	   0.0593&	 	 0.0592&	  6.59E-05&	 0.0158&	0.0013&	 0.0144&    0.07516&	11.657&	 0.81\%\\\hline

  \end{tabular}
\end{table*}

\subsection{Results}
\begin{figure}
\centering
   \begin{subfigure}[b]{0.47\textwidth}
   \includegraphics[height=1.5in, width=1\textwidth]{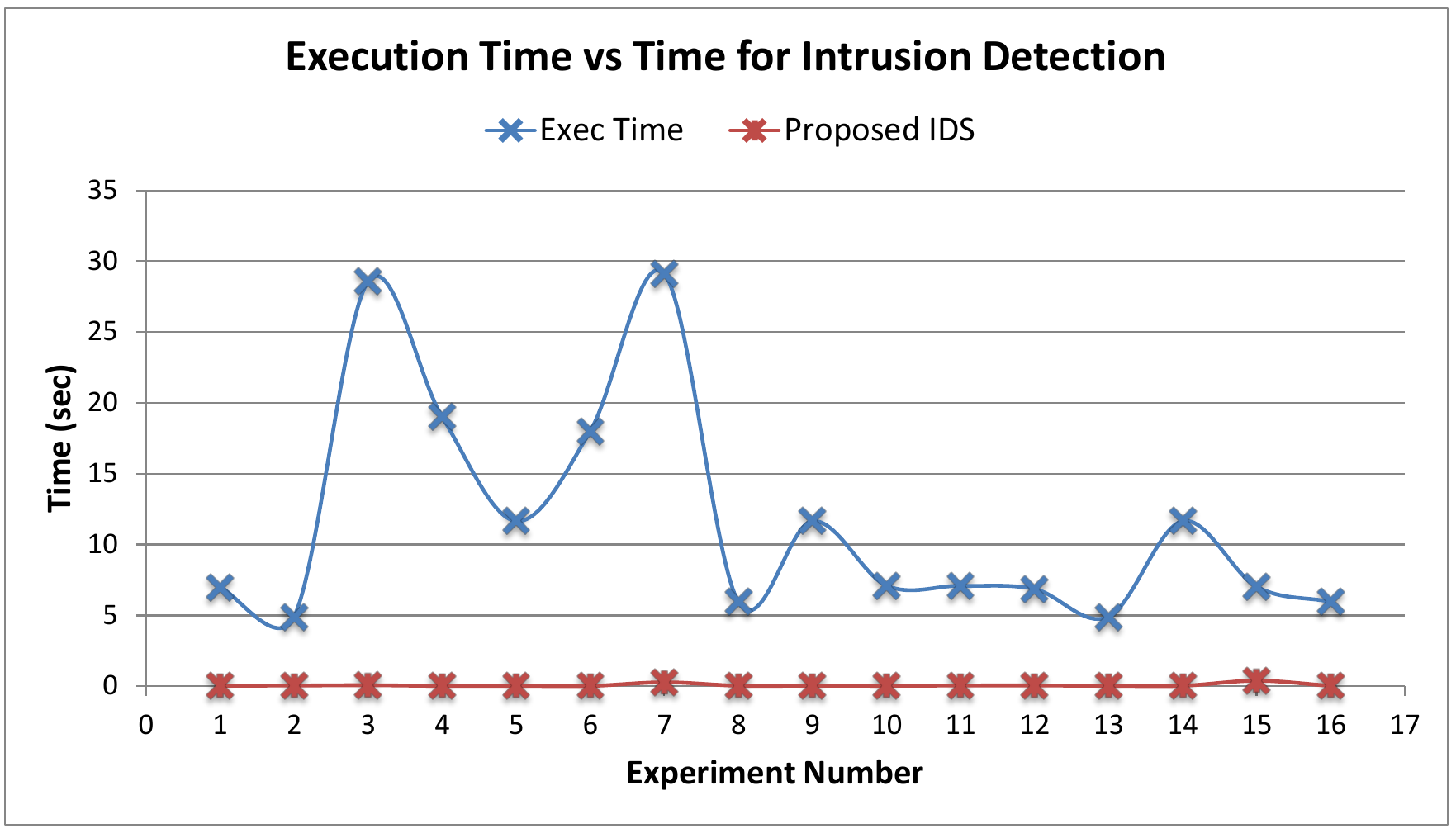}
   \caption{•}
   \label{fig_r1} 
\end{subfigure}
~
\begin{subfigure}[b]{0.47\textwidth}
   \includegraphics[height=1.5in, width=1\textwidth]{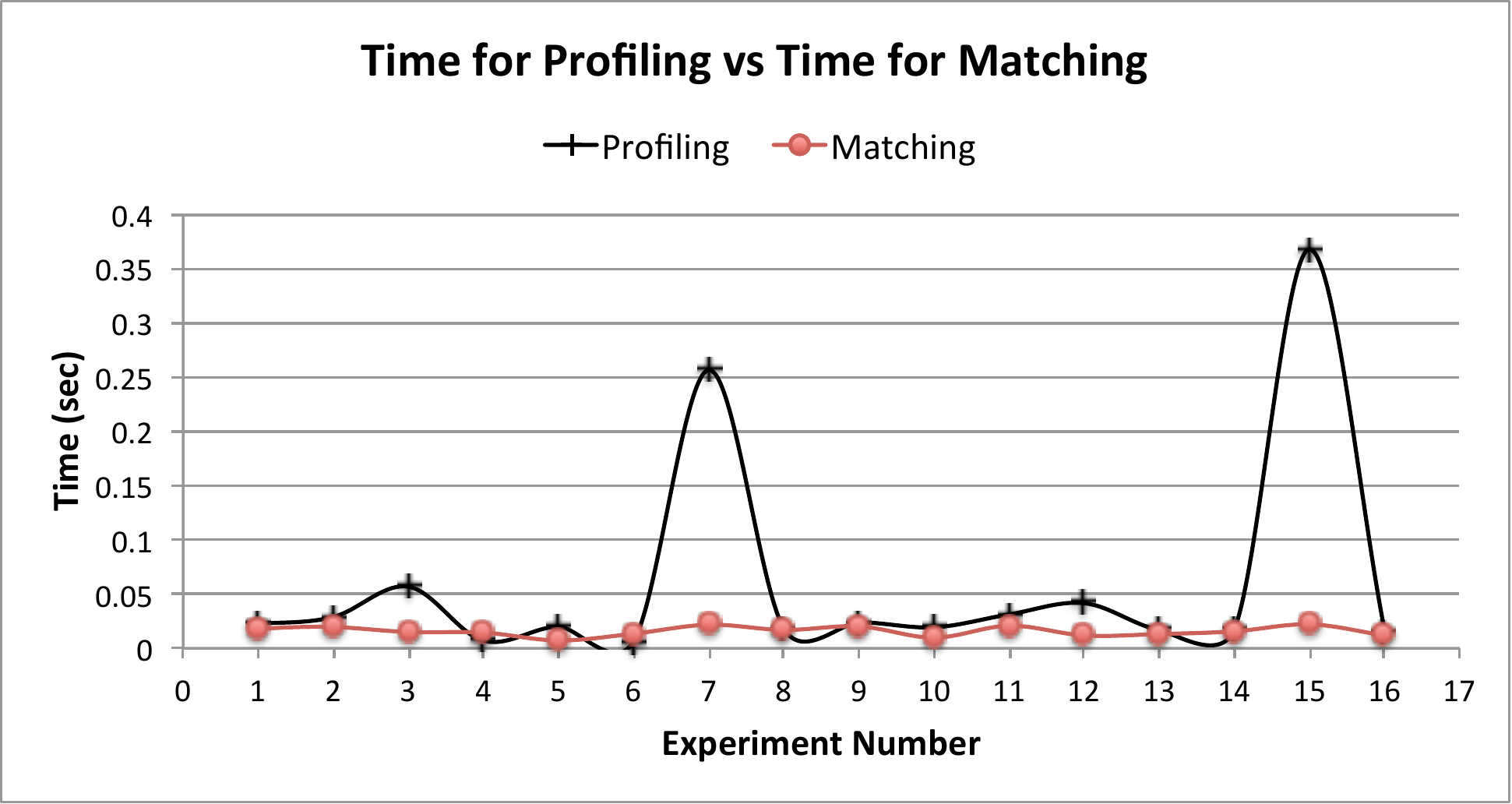}
   \caption{•}
   \label{fig_r2}
\end{subfigure}

\caption[Time Comparisons]{ A time comparison between (a) Proposed IDS technique and run-time for map-reduce examples. (b) Profiling and matching methods of the proposed IDS technique.}
\end{figure}

Table \ref{table_hadoop_values}, Figures \ref{fig_r1} and \ref{fig_r2} show the results of our experiments. Figure \ref{fig_r1} shows the comparison between the time taken to run the hadoop map-reduce examples on a big data cluster and the time taken to run the proposed intrusion detection technique. The execution times for some examples (represented by * in table \ref{table_hadoop_values}) are inconsistent among multiple runs. We can notice from table \ref{table_hadoop_values} that only 0.81\% of time taken to execute an example is needed to analyze it for intrusion detection. The time needed to run the proposed detection technique includes (a) time taken to create CFG for the main method from the class file; (b) time taken to extract MST set from CFG; (c) time taken to hash the MSTs and encrypt them and; (d) time taken to check for similarity among duplicate processes by comparing the program signatures. All of these values can be found in table \ref{table_hadoop_values}. The last row of this table gives the average values. It can be noticed from Figure \ref{fig_r2} that the time required by the proposed technique is influenced by the profiling method trying to extract MSAs from CFG, particularly when there are more than one MSAs for a CFG. Though the matching method performance is directly proportional to the square of the size of the number of edge-disjoint MSAs in a CFG i.e.\ $O(n^2)$ worst case complexity, we observed that it is rare to have more than a couple of edge-disjoint MSAs in a CFG because of the sparse nature of CFG. 

%

\section{Conclusion and Future Work}
In this paper, we introduced a novel approach to detect program level intrusions in big data systems with help of control flow analysis. The main idea is to use the replication property of big data systems and check for coherence in program behavior among replica datanodes. Behavior of a program is modeled by extracting a MSA set representation of its CFG. Similarity check among duplicate programs is performed by a complete matching among hashed sets of MSAs. Experiments were conducted on real-world hadoop map-reduce examples and it is observed that the proposed technique takes only 0.8\% of execution time to identify intrusions. The naturally sparse nature of CFGs helps in achieving this low overhead. For future work, we would like to explore graph string matching and compare the proposed matching method (step2) with other graph isomorphism techniques.


\begin{thebibliography}{1}
	\bibitem{:mapred}
	Dean, Jeffrey, and Sanjay Ghemawat. "MapReduce: simplified data processing on large clusters." Communications of the ACM 51.1 (2008): 107-113.
	\bibitem{:hadoop}
	White, Tom. "Hadoop: The definitive guide." O'Reilly Media, Inc., 2012.
	\bibitem{:spark}
	Zaharia, Matei, et al. "Spark: cluster computing with working sets." Proceedings of the 2nd USENIX conference on Hot topics in cloud computing. 2010.
	\bibitem{:kerberos} 
	O’Malley, Owen. "Integrating kerberos into apache hadoop." Kerberos Conference. 2010.	
	\bibitem{:santosh}
	Aditham, Santosh, and Nagarajan Ranganathan. "A novel framework for mitigating insider attacks in big data systems." Big Data (Big Data), 2015 IEEE International Conference on. IEEE, 2015.
	\bibitem{:cids}
	Tan, Zhiyuan, et al. "Enhancing big data security with collaborative intrusion detection." Cloud Computing, IEEE 1.3 (2014): 27-33.
	\bibitem{:cfg-match1}
	Bruschi, Danilo, Lorenzo Martignoni, and Mattia Monga. "Detecting self-mutating malware using control-flow graph matching." Detection of Intrusions and Malware \& Vulnerability Assessment. Springer Berlin Heidelberg, 2006. 129-143.
	\bibitem{:cfg-match2}
	Nagarajan, Vijay, et al. "Matching control flow of program versions." Software Maintenance, 2007. ICSM 2007. IEEE International Conference on. IEEE, 2007.
	\bibitem{:graph-compare-exe}
	Dullien, Thomas, and Rolf Rolles. "Graph-based comparison of executable objects (english version)." SSTIC 5 (2005): 1-3.
	\bibitem{:ligatti}
	Abadi, Martín, et al. "Control-flow integrity principles, implementations, and applications." ACM Transactions on Information and System Security (TISSEC) 13.1 (2009): 4.
	\bibitem{:cfg-correct}
	Amighi, Afshin, et al. "Provably correct control flow graphs from Java bytecode programs with exceptions." International Journal on Software Tools for Technology Transfer (2015): 1-32.
	\bibitem{:cfg-reduce}
	Gold, Robert. "Reductions of Control Flow Graphs." World Academy of Science, Engineering and Technology, International Journal of Computer, Electrical, Automation, Control and Information Engineering 8.3 (2014): 417-424.
	\bibitem{:edmonds1}
	Gabow, Harold N., et al. "Efficient algorithms for finding minimum spanning trees in undirected and directed graphs." Combinatorica 6.2 (1986): 109-122.
	\bibitem{:edmonds2}
	Uno, Takeaki. An algorithm for enumerating all directed spanning trees in a directed graph. Springer Berlin Heidelberg, 1996.
	\bibitem{:edmonds3}
	J. Edmonds, Optimum branchings, J. Res. Natl. Bur. Standards 71B (1967), 233–240.
	\bibitem{:edit-distance}
	Bunke, Horst. "On a relation between graph edit distance and maximum common subgraph." Pattern Recognition Letters 18.8 (1997): 689-694.
	\bibitem{:traversal}
	Sharir, Micha. "A strong-connectivity algorithm and its applications in data flow analysis." Computers \& Mathematics with Applications 7.1 (1981): 67-72.
	\bibitem{:tarjan1}
	Georgiadis, Loukas, Robert Endre Tarjan, and Renato Fonseca F. Werneck. "Finding Dominators in Practice." J. Graph Algorithms Appl. 10.1 (2006): 69-94.
	\bibitem{:tarjan2}
	Tarjan, Robert E., and Mihalis Yannakakis. "Simple linear-time algorithms to test chordality of graphs, test acyclicity of hypergraphs, and selectively reduce acyclic hypergraphs." SIAM Journal on computing 13.3 (1984): 566-579.
	\bibitem{:ids-book}
	Pathan, Al-Sakib Khan, ed. The state of the art in intrusion prevention and detection. CRC press, 2014.
	\bibitem{:static-ids}
	Wagner, David, and Drew Dean. "Intrusion detection via static analysis." Security and Privacy, 2001. S\&P 2001. Proceedings. 2001 IEEE Symposium on. IEEE, 2001.
	\bibitem{:hdfs-tracing}
	Wang, William. End-to-end Tracing in HDFS. Diss. Carnegie Mellon University Pittsburgh, PA, 2011.
	\bibitem{:pivot-tracing}
	Mace, Jonathan, Ryan Roelke, and Rodrigo Fonseca. "Pivot tracing: dynamic causal monitoring for distributed systems." Proceedings of the 25th Symposium on Operating Systems Principles. ACM, 2015.
	\bibitem{:algos-graphsim}
	Koutra, Danai, et al. Algorithms for graph similarity and subgraph matching. Technical Report of Carnegie-Mellon-University, 2011.
	\bibitem{:isomorphism}
	Cordella, Luigi P., et al. "A (sub) graph isomorphism algorithm for matching large graphs." Pattern Analysis and Machine Intelligence, IEEE Transactions on 26.10 (2004): 1367-1372.
	\bibitem{:cayley}
	Shor, Peter W. "A new proof of Cayley's formula for counting labeled trees." Journal of Combinatorial Theory, Series A 71.1 (1995): 154-158.
	\bibitem{:amazon}
	Amazon, E. C. "Amazon elastic compute cloud (Amazon EC2)." Amazon Elastic Compute Cloud (Amazon EC2) (2010).
	\bibitem{:sage}
	Sage Mathematics Software (Version 4.0), The Sage Developers, 2016, http://www.sagemath.org.
	\bibitem{:gml}
	Brandes, Ulrik et al. “Graph Markup Language (GraphML).” CRC (2013). 
	\bibitem{:gv}
	Emden R. Gansner and Stephen C. North. "An open graph visualization system and its applications to software engineering." SOFTWARE - PRACTICE AND EXPERIENCE 30.11 (2000): 1203-1233.
	\bibitem{:nx}
	Aric A. Hagberg, Daniel A. Schult and Pieter J. Swart, “Exploring network structure, dynamics, and function using NetworkX”, in Proceedings of the 7th Python in Science Conference (SciPy2008), Gäel Varoquaux, Travis Vaught, and Jarrod Millman (Eds), (Pasadena, CA USA), pp. 11–15, Aug 2008 
	\bibitem{:cfgf}
	Alekseev, Sergej, Peter Palaga, and Sebastian Reschke. "Bytecode Visualizer." Control Flow Graph Factory. N.p., 2008. Web. 24 Mar. 2016.
\end{thebibliography}
\end{document}